# Unusual rotating magnetocaloric effect in the hexagonal ErMnO$_3$ single crystal

M. Balli[1,2] *, S. Jandl [1,2], P. Fournier [1,2,3], J. Vermette [4], D. Z. Dimitrov [5,6,7]

[1] Institut Quantique, Université de Sherbrooke, J1K 2R1, QC, Canada.

[2] Regroupement québécois sur les matériaux de pointe, Département de physique, Université de Sherbrooke, J1K 2R1, QC, Canada.

[3] Canadian Institute for Advanced Research, Toronto, Ontario M5G 1Z8, Canada.

[4] Département de chimie, Université de Sherbrooke, J1K 2R1, QC, Canada.

[5] Institute of Solid State Physics, Bulgarian Academy of Science, Sofia 1184, Bulgaria.

[6] Institute of Optical Materials and Technologies, Bulgarian Academy of Sciences, Sofia 1113, Bulgaria.

[7] Department of Electrophysics, National Chiao Tung University, Hsinchu 30010, Taiwan.

**Abstract**

It is known that orthorhombic RMnO$_3$ multiferroics (R = magnetic rare earth) with low symmetry exhibit a large rotating magnetocaloric effect because of their strong magnetocrystalline anisotropy. In this paper, we demonstrate that the hexagonal ErMnO$_3$ single crystals also unveils a giant rotating magnetocaloric effect that can be obtained by spinning them in constant magnetic fields around their a or b axes. When the ErMnO$_3$ crystal is rotated with the magnetic field initially parallel to the c-axis, the resulting entropy change reaches maximum values of 7, 17 and 20 J/kg K under a constant magnetic field of 2, 5 and 7 T, respectively. These values are comparable or even larger than those shown by some of the best orthorhombic phases. More interestingly, the generated anisotropic thermal effect is about three times larger than that exhibited by the hexagonal HoMnO$_3$ single crystal. The enhancement of the rotating magnetocaloric effect in the hexagonal ErMnO$_3$ compound arises from the unique features of Er$^{3+}$ magnetic sublattice. In fact, the Er$^{3+}$ magnetic moments located at 2a sites experience a first-order metamagnetic transition close to 3 K along the c-axis resulting in a peaked magnetocaloric effect over a narrower temperature range. In contrast, the "paramagnetic" behaviour of Er$^{3+}$ magnetic moments within the ab-plane, produces a larger magnetocaloric effect over a wider temperature range. Therefore, the magnetocaloric effect anisotropy is maximized between the c and the ab-directions, leading to a giant rotating magnetocaloric effect.

*Mohamed.balli@usherbrooke.ca

**I. Introduction**

Based on the well-known magnetocaloric effect (MCE), magnetic cooling is a trending technology that continues to attract a worldwide interest due to its potential high thermodynamic efficiency as well as its eco-friendly character [1-8]. The implementation of this emergent technology in our daily life would enable to fully suppress the harmful synthetic refrigerants such as fluorinated fluids, usually present in standard refrigerators and air-conditioners [1-8], thus allowing to meet the requirements of several treaties that were universally adopted by the international community aiming to reduce the utilization of CFCs, HCFCs and HFCs gases and, green house gases (GHG) emissions [1]. However, the search for advanced magnetocaloric materials with outstanding thermal, chemical and mechanical properties is necessary for the transfer of magnetic refrigeration technology towards the market place. In this context, a large MCE has been pointed out in a wide variety of magnetocaloric materials including both intermetallics and oxides [1-8]. Some of them such as $LaFe_{13-x}Si_x$ compounds, $Fe_2P$-type materials and $ABO_3$-based oxides have been successfully tested in room-temperature devices unveiling the bright future of magnetic refrigeration [1]. On the other hand, magnetic materials with excellent caloric effects at the temperature range below 30 K are of great importance in several low-temperature applications such as space industry, scientific facilities and, hydrogen and helium liquefiers. Particularly, the production and utilization of liquified hydrogen is expected to markedly increase in the forthcoming years as an alternative source of energy [8]. To deal with the increasing demand for efficient solid state-refrigerants operating at very low temperatures, worldwide research works have been conducted leading to quite interesting systems such as $DyGdAl_5O_{12}$ garnets [9, 10], $RAl_2$ intermetallics (R = rare earth) [5], $RMn_2O_5$ [11, 12] multiferroics and $RMO_3$ oxides (M = transition metal) [13-27].

Particularly, the $RMnO_3$ compounds are still attracting a growing interest, because of their excellent magnetocaloric properties as well as their potential implementation in spintronic devices [13-27, 28-40]. Also, these oxides unveil degrees of freedom other than magnetization such as electric polarization, charge and volume change enabling additional potential caloric effects. Their structural and physical properties strongly depend on the rare-earth size [28-40]. Usually when the ionic radius of the rare-earth is lower than that of Dy, $RMnO_3$ crystallizes in a hexagonal structure with P63cm space group while for larger ionic radius ($r_R > R_{Dy}$), the crystalline structure transforms into an orthorhombic symmetry with Pbnm space group. On the other hand, the ferroelectricity origin as well as the strength of the magnetoelectric coupling in the $RMnO_3$ compounds highly depend on their crystallographic structure. Hence, the hexagonal $RMnO_3$ show a weak magnetoelectric coupling since the ferroelectricity arises from structural deformations. In contrast, the appearance of electric polarization in orthorhombic RMnO$_3$ originates from the magnetic frustration, leading accordingly to a strong coupling between magnetic and electric ordering parameters [28-40]. More recently, it was reported that the gigantic magnetocrystalline anisotropy particularly shown by orthorhombic RMnO$_3$ compounds leads to a giant rotating magnetocaloric effect (RMCE) that can be simply generated by spinning their crystals in constant magnetic fields, instead of the standard magnetization-demagnetization process [27]. Such effect would enable more efficient and compact magnetic cooling machines with simplified designs.

In this paper we mainly focus on the magnetic and magnetocaloric properties of the hexagonal ErMnO$_3$ single crystal (h-ErMnO$_3$), aiming to get more insight about its real potential in low-temperature magnetic cooling. In h-ErMnO$_3$, the competition between different magnetic exchange interactions involving the Mn and Er sublattices gives rise to the appearance of several magnetic phase transitions over a temperature range going from 1000 K down to the liquid helium temperature [41-50]. Therefore, a paraelectric to ferroelectric transition is observed at about 833 K, while the Mn magnetic moments order antiferromagnetically close to 80 K. The Er magnetic moments usually undergo ordering transitions below 10 K. It is worth noting that the Mn magnetic moments are strongly frustrated in the-ab plane within a triangular structure, making marginal their contribution to the total magnetization even under intense magnetic fields [41-50]. For this purpose, our present study of the magnetocaloric properties of h-ErMnO$_3$ single crystals is motivated mainly by the large anisotropy expected in the temperature range around 10 K, where the ordering state of Er magnetic moments can be easily manipulated under relatively low driving magnetic fields. Moreover, the choice of single crystals instead of polycrystalline samples will allow us to avoid grain boundary effects and structural and stoichiometric inhomogeneity. Additionally, the single crystals approach enables us to explore the strength of the MCE anisotropy in h-ErMnO$_3$ and accordingly, the resulting rotating magnetocaloric effect.

## II. Experimental

h-ErMnO$_3$ single crystals were grown using the high temperature solution method with the help of PbF$_2$/PbO/B$_2$O$_3$ flux [51]. The ErMnO$_3$ polycrystalline samples were prepared by the standard solid-state reaction method using stoichiometric quantities of Er$_2$O$_3$ and MnO$_2$ precursors annealed at 1120 °C for 24 h. The flux was then mixed with ErMnO$_3$ polycrystals in a 7:1 proportion and subjected to heat treatment in a platinum crucible at 1250 °C for 48 h. The growth process was achieved by cooling slowly the final product down to 1000 °C at a rate of 5 °C/h [51]. The symmetry of the crystals and their crystalline quality were checked using Raman scattering.

Micro-Raman spectra were collected for two polarizer configurations [b(cc)b and c(ab)c] using a Labram-800 spectrometer equipped with a microscope, a He-Ne laser and a nitrogen-cooled charge coupled device (CCD) detector. c and b represent the incident and scattered propagation directions, while (cc) and (ab) are the polarizer orientations for the incident and scattered beams. Careful investigations of Raman-active modes reported in Figure 1 unveils that like the reference material YMnO$_3$, h-ErMnO$_3$ crystals form in a high-quality hexagonal structure with $P6_3cm$ space group (Fig. 1 right) where the Er$^{3+}$ ions occupy two inequivalent lattice sites, one with C$_{3v}$ symmetry defined as site 2a, and the other one with C$_3$ symmetry defined as site 2b. The magnetization measurements were carried out using a superconducting quantum interference device (SQUID) magnetometer from Quantum Design (MPMS XL7). Aiming to minimize the demagnetization effect, before each measurement we ensure that the target crystallographic direction (ab or c) is along the largest dimension of the selected h-ErMnO$_3$ crystal (roughly 0.5 mm*0.5 mm*5 mm).

III. **Results and discussion**

In Fig.2-a, we report the 0.1 T-zero-field cooled (ZFC) and field-cooled (FC) magnetization as a function of temperature measured following the c-axis and the ab-plane. As can be seen for both directions, the thermomagnetic curves exhibit a nearly reversible behaviour unveiling the absence of thermal hysteretic losses. As the temperature decreases, the magnetization is markedly enhanced at very low temperatures, which is mainly caused by the ordering of Er$^{3+}$ magnetic moments. The linear fitting of the 0.1 T-inverse susceptibility (1/χ) shown in the inset of Figure 2-a using the Curie-Weiss law χ$^{-1}$=(T-T$_\theta$)/C (C is the Curie-Weiss constant) leads to paramagnetic Curie temperatures (T$_\Theta$) of about -3 and -63 K for the ab and c-orientations, respectively. The strong negative value of T$_\Theta$ reflects the presence of dominant antiferromagnetic interactions in the hexagonal ErMnO$_3$ along the c-axis. In contrast, its weak negative value for the field applied along the ab-plane indicates a weak antiferromagnetic ordering and/or a paramagnetic behavior of Er$^{3+}$ magnetic moments. The deduced effective magnetic moment from the linear interpolation of 1/χ at high temperatures was found to be 10.14 μ$_B$ for the c-axis and 10.39 μ$_B$ for the ab-plane, being in perfect accord with the theoretical value given by $\mu_{eff} = \sqrt{(\mu_{eff}(Er^+))^2 + (\mu_{eff}(Mn^{3+}))^2} = 10.75 \mu_B$: here, the effective magnetic moments of Er$^{3+}$ and Mn$^{3+}$ ions [52] were taken as 9.58 μ$_B$ and 4,889 μ$_B$, respectively.

It is worth noting that with increasing magnetic fields parallel to the c-axis, a sudden change in the magnetization at about 2 K can be clearly seen from the thermomagnetic curves reported in Figure 2-b. In addition, this transition zone slightly moves to high temperatures under magnetic fields above 1 T. The observed behavior contrasts markedly to that shown by the magnetization along the ab-plane. This can be mainly explained by the fact that the $Er^{3+}$ magnetic moments undergo a field-induced metamagnetic transition from the antiferromagnetic to the ferromagnetic state. On the other hand, the anomaly at ~80 K associated with the antiferromagnetic ordering of $Mn^{3+}$ magnetic moments is not clearly visible from the reported thermomagnetic curves in Figure 2-a. The absence of $Mn^{3+}$ critical point can be essentially attributed to the frustrated triangular magnetic structure of the $Mn^{3+}$ magnetic moments within the ab-plane and the dominance of the magnetic moment of $Er^{3+}$. In the multiferroic h-$ErMnO_3$, the $Mn^{3+}$ ions organize in layers of triangular sub-units stacked along the c-axis (Fig. 1), while their related moments form an angle of 120 ° with one another, leading to a negligible net magnetization [50]. Consequently, the magnetic features involving the $Mn^{3+}$ sub-lattice are masked by the large magnetization of the $Er^{3+}$ sub-lattice. Figure 3-a displays the magnetization as a function of magnetic field along the ab-plane and the c-axis at 2 K. As can be seen, the h-$ErMnO_3$ single crystal unveils a large magnetocrystalline anisotropy which is a common feature of frustrated $RMnO_3$ manganites [13-40]. However, in comparison to orthorhombic $RMnO_3$ phases [13-40], this magnetic anisotropy remains less pronounced because of the high crystallographic symmetry. According to data of Figures 2-a and 3-a, the magnetization behavior clearly indicates that the ab-plane corresponds to the easy-orientation, while the c-axis looks like a hard-direction. Such behavior is intriguing since the preferred orientation of $Er^{3+}$ magnetic moments is along the c-axis. In this case, the $Mn^{3+}$ moments create a molecular magnetic field in this direction [45, 47], which is also the case of the hexagonal $DyMnO_3$ single crystal. In this context, it is worth to underline the work by Skumryev et al [44] in which the magnetic anisotropy of hexagonal $RMnO_3$ manganites was rather attributed to the quadrupolar charge distribution of the 4f shells of the rare-earth elements.

Along the "easy-plane" ab, the magnetization increases with magnetic field following a paramagnetic trend without any steplike transitions, to finally reach a large value under high fields. The magnetization at 7 T applied within the ab-plane is about 160 $Am^2/kg$ (7.74 $\mu_B$/f.u.), which is more than 85% of the $Er^{3+}$ free magnetic moment (9 $\mu_B$) [52] and 60 % of the total magnetization in the case where the magnetic moments of both, $Er^{3+}$ (9 $\mu_B$) and $Mn^{3+}$ (4 $\mu_B$) [52] sublattices are fully aligned in the direction of the magnetic field. Keeping in mind the marginal contribution of the Mn sublattice to the total magnetization, this result reveals that the $Er^{3+}$ magnetic moments can be completely aligned when applying sufficiently high magnetic fields along the ab-plane. In contrast, the h-$ErMnO_3$ crystal shows distinguishing features along the c-axis. It behaves like a ferromagnetic material under very low magnetic fields (< 0.1 T). This "ferromagnetic" volume vanishes at about 2.8 K (Fig. 3-b). Meier et al [41] attributed this pronounced response under small magnetic fields to the formation of a ferrimagnetic multidomain state of $Er^{3+}$ magnetic moments. However, as this remnant magnetization is negligible, the observed polarization is most probably associated with the appearance of a weak ferromagnetism due to spin canting, instead of a normal ferromagnetism or ferrimagnetism [47]. For magnetic fields higher than 0.1 T applied along the c-axis (Fig. 3-b), the magnetization increases almost linearly followed by a steplike enhancement after overpassing a critical field of roughly 0.8 T. For magnetic fields larger than 1 T (Fig. 3-a), the magnetization continues to increase slightly without any tendency to saturate even under high magnetic fields. It reaches 40 $Am^2/kg$ under a magnetic field of 7 T, being only 21.5 % of that shown by the $Er^{3+}$ free ion. On the other hand, the h-$ErMnO_3$ crystal shows a large magnetic anisotropy when compared to other hexagonal $RMnO_3$ phases such as $HoMnO_3$ [16] and $DyMnO_3$ [26] single crystals. When the magnetic field is changed from the easy-direction to the hard-direction, the magnetization at 7 T is reduced by 75 % for h-$ErMnO_3$ and only 46 % for h-$HoMnO_3$ [16].

As explained above, it is highly challenging to manipulate the magnetic order of $Mn^{3+}$ sublattice even under intense magnetic fields, since the $Mn^{3+}$ magnetic moments are strongly coupled in the ab-plane. This low magnetic response of the $Mn^{3+}$ sublattice implies that the magnetocaloric properties of h-$ErMnO_3$ single crystals around the magnetic ordering point are involving mainly the $Er^{3+}$ sublattice. The MCE was initially calculated in term of the isothermal entropy change by numerically integrating the reported magnetic isotherms in Figure 4, using the well-known Maxwell relation [1-7] given by:

$$\Delta S (T, 0-B) = \int_0^B \left(\frac{\partial M}{\partial T}\right)_{B'} dB' \qquad (1)$$

In the absence of hysteretic effects [53], Eq. (1) enables us a fast and accurate characterization of magnetocaloric materials. It is worth noting that the utilization of Maxwell equation and magnetization data to evaluate the entropy change was a subject of controversy in the past [1, 53-55]. However, this mainly concerned materials that present large hysteretic effects resulting in phase-separated states which is not the case of the here studied compound. In fact, the h-$ErMnO_3$ crystal unveils almost zero hysteresis along the ab and c-directions. On the other hand, magnetization data were successfully used in the past to calculate the entropy change in similar materials ($RMnO_3$) showing both first and second order magnetic phase transitions [13-27].

Figure 5-a, b presents the deduced -$\Delta S$ as a function of temperature under some representative magnetic fields applied along the ab-plane and the c-axis. Because of the magnetic frustration, the entropy change presents

a large anisotropy. Following the magnetization trend, giant values for the entropy change can be obtained along the ab-plane. In magnetic fields changing from 0 to 2 T, 0 to 5 T, and 0 to 7 T, -ΔS reaches maximum values of about 12.4, 20.5, and 22.7 J/kg K close to the ordering temperature of $Er^{3+}$ magnetic moments (~10 K), respectively. Although the h-ErMnO$_3$ crystal contains a rare-earth element with low magnetic moment as compared to $Dy^{3+}$ and $Ho^{3+}$ ions (10 $\mu_B$) [52], its MCE in terms of -ΔS largely exceeds that shown by other hexagonal RMnO$_3$ manganites such as h-DyMnO$_3$ and h-HoMnO$_3$ compounds [16, 26]. In similar magnetic fields applied within the same plane at similar temperature range, the maximum values of –ΔS are evaluated to be 3.86, 13.9 and 18.7 J/kg K for h-HoMnO$_3$ [16] and, 8, 15 and 19 J/kg K for h-DyMnO$_3$ [26], respectively. The marked difference between entropy changes shown by h-RMnO$_3$ (R = Er, Ho, Dy) crystals are mainly attributed to the distinguished magnetic features of their rare-earth elements. Such features are usually driven by the 4f-4f and 3d-4f magnetic interactions that significantly differ from one h-RMnO$_3$ to the other. For more details, we refer the interested reader to Refs. 27-40.

Because of the strong magnetocrystalline anisotropy, the entropy change following the c-axis was found to be much lower, with respect to the ab-plane. When changing the applied magnetic field along the c-direction from 0 to 2 T, 0 to 5 T and 0 to 7 T, -ΔS reaches maximum values of about 6, 9.5 and 11.4 J/kg K, respectively. These values compare well with those obtained for h-HoMnO$_3$ [16] and h-DyMnO$_3$ [26] under relatively high magnetic fields applied along the same crystallographic axis. However, to get more insight about the mechanism behind the magnetocaloric effect along the c-axis, the contribution of the metamagnetic-like transition was also investigated. Its associated entropy change can be well evaluated by using the Clausius-Clapeyron equation given by [53, 56]:

$$\Delta S_{meta} = -\Delta M \frac{dB_C}{dT} = -\Delta M \left(\frac{dT_T}{dB}\right)^{-1} \qquad (2)$$

where $B_C$ and $T_T$ are the critical field and the transition temperature, respectively. ΔM is the magnetization jump. In contrast to the Maxwell relation, the above equation enables us to directly link ΔS with the sudden change in the magnetization. The critical field which is usually defined as the inflection point shown by isotherms within the metamagnetic zone is reported in Figure 6-a as a function of temperature. When increasing temperature, it evolves almost linearly with a positive rate of about 0.4 T/K. The temperature dependence of -ΔS$_{meta}$ is plotted in Figure 6-b. As shown, the MCE at very low temperatures originates mainly from the $Er^{3+}$ magnetic moments on the 2a sites that experience a field-induced magnetic transition from an antiferromagnetic state toward a ferromagnetic order [41, 45, 47], while those on the 4b sites contribute for little. For example, the application of a magnetic field of 2 T parallel to the c-axis at 1.9 K results in a full entropy change of 5.33 J/kg K, while the contribution of the metamagnetic region is evaluated to be 5 J/kg K. Even though the magnetic field is increased by 5 T (from 2 to 7 T) the entropy change is enhanced only by 2 J/kg K which can be attributed to the ordering of the magnetic moments on the 4b sites. These findings also underline the strong coupling between the $Mn^{3+}$ and the $Er^{3+}$ moments on the 4b sites that forbids their easy orientation or ordering in the presence of external magnetic fields, giving then rise to moderate levels of MCEs. However, as can be seen in Figure 6-b the MCE associated with the transitional zone weakens while the temperature increases which is due to the disordering of the magnetic moments at the 2a sites. At high temperatures, the MCE is expected to originate from both 4b and 2a magnetic moments.

The driving forces behind the MCE in h-ErMnO$_3$ along the c-axis can be well understood when looking carefully at the complex magnetic structure of the Er sublattice. According to early works [45], it was found that the 120 ° antiferromagnetic ordering of the $Mn^{3+}$ magnetic moments at the Néel temperature (80 K) triggers the magnetic order of the $Er^{3+}$ magnetic moments occupying the 4b sites, while those located at the 2a sites remain disordered. This is mainly due to the strong coupling between the $Mn^{3+}$ ($3d^4$) and $Er^{3+}$($4f^{11}$) 4b magnetic properties. By using inelastic neutron scattering, terahertz and far infrared spectroscopies as well as some theoretical models, Chaix et al [45] have unveiled a distinct hybridization between the magnetically coupled Mn and Er (4b) via a crystalline field transition. In contrast, there is no coupling between the $Er^{3+}$(2a) and $Mn^{3+}$ sublattices [45]. Because of the mean field created by the Mn magnetic moments, it was found that the Er magnetic moments at the 4b sites are antiferromagnetically polarized along the c-axis [45]. According to Meir et al [41], a ferromagnetic ordering of $Er^{3+}$ magnetic moments at the 2a site occurs below 10 K following the c-axis. These new 2a established exchange interactions result in the reorientation of both $Er^{3+}$ (4b) and $Mn^{3+}$ magnetic moments at about 7 and 2.5 K, respectively. Meir et al [41] also claimed that at very low temperatures the $Er^{3+}$ (4b) and $Er^{3+}$ (2b) sites maintain a ferrimagnetic ground state. However, the reported magnetic structure in Ref .41 fails to explain the multiple magnetic phase transitions experienced by h-ErMnO$_3$ under intense magnetic fields [47]. In a recent report by Song et al [47], it was found that in addition to the metamagnetic transition that occurs in an external magnetic field close to 1 T, h-ErMnO$_3$ exhibits other magnetic phase transitions when subjected to 12 and 28 T [47]. According to these reported data [47], it seems that the $Er^{3+}$ magnetic moments located at 2a sites are rather antiferromagnetically ordered along the c-axis. Furthermore, under the effect of an external magnetic field of about 1 T, they present a metamagnetic like transition to reach a ferromagnetic state (Fig.3), leading accordingly to high levels of MCE. The $Er^{3+}$ magnetic moments on the 4b sites exhibit a spin reorientation under 12 T before being fully oriented parallel to the c-axis at 28 T [47]. In contrast to the magnetic moments on the 2a sites, such spins ordering is accompanied by a small change in the magnetization giving rise to negligible thermal effects.

The strong coupling between the Mn and 4f (4b) magnetic sublattices seems to be supported by Raman scattering data. Figure 7 presents two phonon frequency shifts as a function of temperature for h-ErMnO$_3$ and YMnO$_3$ compounds. The first panel (a) is related to the phonon involving the Mn and the basal oxygen ions (O2,3) oscillating in the ab-plane (out-of-phase). For both materials, the frequency shift for this mode follows the typical anharmonic phonon temperature dependence (solid line) until an extra hardening occurs below 80 K (T$_N$). This particular atomic motion tends to deform the Mn$^{3+}$ magnetic arrangement in the ab-plane resulting in an additional contribution to the constant force trough spin-phonon interaction [42]. This extra hardening has been observed in all the hexagonal manganites [57]. The second panel (b) represents the temperature dependence of the phonon related to the apical oxygen ions (O1,2) oscillating along the c-axis. This mode involves atomic motions that are usually independent of the Mn$^{3+}$ spin frustration when the rare-earth element has no magnetic moment [57] such as YMnO$_3$. Otherwise, a phonon extra hardening is observed below T$_N$ as can be seen for the h-ErMnO$_3$ crystal. This underlines an anisotropic superexchange magnetic interaction between the Mn$^{3+}$ planes and the rare-earth ions through the apical oxygen ions (see Fig. 1 with orange shaded line). This type of magnetic interactions was found to be responsible for the appearance of magnetoelectric effects in the HoMnO$_3$ manganite [58].

It is worth noting that the theoretical limit that can be reached by the entropy change in h-ErMnO$_3$ is about $\Delta S_{Limit} = R*Ln(2J+1) = 81$ J/kg K, with R is the universal gas constant and J is the angular quantum momentum. The latter can be determined (effective value) from the magnetization saturation (g*J*µ$_B$ = 13 µ$_B$) assuming the Landé factor g equal to 2. On the other hand, the obtained entropy change in the field variation of 7 T applied within the ab-plane mostly arising from the Er sublattice, is only 27 % of $\Delta S_{Limit}$ unveiling the hidden magnetocaloric potential of h-ErMnO$_3$ single crystals. To attain the theoretical limit, it is necessary to orient the magnetic moments of both Er$^{3+}$ (9 µ$_B$) and Mn$^{3+}$ (4 µ$_B$) from a perfect paramagnetic to a complete ordered state. This can be for example carried out under intense magnetic fields. However, in the case of h-ErMnO$_3$, this seems to be highly challenging from a practical point of view since unusual magnetic fields far above 50 T are required [47]. In a recent work by Song et al [47], it was shown that a magnetic field of 50 T is necessary to fully align the Er$^{3+}$ magnetic moments, while the triangular frustration of Mn$^{3+}$ magnetic moments (negligible magnetization) remains stronger even under the application of this "colossal" magnetic field. In order to break up the 120 ° coupling and allowing the Mn sublattice to contribute to the MCE, new routes other than applying large magnetic fields must be explored. For example, the manipulation of the 4f-3d exchange interactions through chemical doping and/or external excitations such as pressure and electric field would be interesting ways to investigate.

The refrigerant capacity (RC) [59] is an important "practical" parameter that enables us to probe the applicability of magnetocaloric materials. This is because the RC takes into consideration both the entropy change magnitude and the working temperature range. It is usually given by

$$RC = \int_{T_C}^{T_H} \Delta S(T) dT \qquad (3)$$

where $T_C$ and $T_H$ are the cold and hot temperatures corresponding to the half maximum of ΔS (T). It is worth noting that in certain cases the RC cannot be used as a reliable parameter for the characterization of magnetocaloric materials. This usually concerns some specific materials that exhibit a weak entropy change over a very large temperature range, giving rise to unreasonable large values for RC. However, this is not the case of the present studied crystal that exhibits a giant entropy change over a wide working temperature window. In fact, sufficiently high MCE levels are necessary to achieve an efficient refrigeration process via the enhancement of heat exchanges between the magnetocaloric refrigerant and the carrier fluid in cooling devices. In this case, the RC could be used as a good figure of merit to identify the potential of magnetocaloric materials in practical applications [60]. For example, it was more recently shown by Niknia et al [60] that the RC linearly scales with the exergetic cooling power of an AMR-magnetic refrigerator. For the h-ErMnO$_3$ compound, the obtained RC in the magnetic field change of 7 T applied within the ab-plane is more than 440 J/kg. This value is comparable to that of the best orthorhombic RMnO$_3$ such as o-DyMnO$_3$ [27] and, much larger than those early reported for hexagonal RMnO$_3$ phases such as h-HoMnO$_3$ and h-DyMnO$_3$ single crystals [16, 26, 27]. Along the c-axis, the h-ErMnO$_3$ reveals a moderate RC of only about 14 % (64 J/kg) of that observed for a field along the ab-plane (for 7 T). On the other hand, in a similar magnetic field, the exhibited RC by h-DyMnO$_3$ and h-HoMnO$_3$ single crystals along their c-axis [16, 26] is about three times larger if compared to that of h-ErMnO$_3$ along the same axis. Such marked differences could be mainly attributed to the specific magnetic features of h-RMnO$_3$ (R = Er, Ho. Dy) crystals as explained above. In addition, the h-ErMnO$_3$ exhibits a narrow magnetocaloric operating temperature range along the c-axis, markedly contrasting with the ab-plane as shown in Figure 8-a. Under 7 T, the difference $T_H$-$T_C$ is lower than 7 K following the c-axis, while it is more than 26 K along the ab-plane. The limited operating magnetocaloric window along the c-axis is mainly attributed to the first order character of the AF-F transition taking place at temperatures below 4 K (see Figs. 2 and 3). In fact, the nature of the magnetic transitions along the ab and c-directions were checked using the Banerjee criterion also known as Arrott plots [61] (Fig. 8-c, d). This approach consists in the analysis of the slope of $M^2$ versus H/M isotherms. Their negative slope (Fig. 8-c) along the c-axis (S-like shape) indicates that the phase transition associated with the ordering of $Er^{3+}$ magnetic moments located at 2a sites (at very low temperatures) is first-order in nature. Also, the recently reported specific heat measurements by Song et al. [47] for h-ErMnO$_3$ along the c-axis present pronounced spikes for magnetic fields higher or equal to 1 T being a signature of first order phase transitions. On the other hand, a rapid change in the magnetization along the c-axis after overpassing a critical field (see Fig3-b), while the latter increases with increasing temperature (Fig.6-a), is another indication for the occurrence of a first order magnetic transition along the c-axis [53]. In contrast, the positive slope of $M^2$ versus H/M along the ab-plane (Fig. 8-d) indicates that the "paramagnetic" ordering of $Er^{3+}$ sublattice is of second order.

It is known that the delivered cooling power by a magnetocaloric device is mainly determined by the magnitude of the entropy change. However, the adiabatic temperature change $\Delta T_{ad}$ is extremely important since it strongly influences the strength of heat exchanges and accordingly the temperature span between its hot and cold sources. As the h-ErMnO$_3$ crystal exhibits a second magnetic order transition along the ab-plane, its adiabatic temperature change was deduced by using the following equation [1, 14]:

$$\Delta T_{ad} = -\frac{T}{C_P}\Delta S \qquad (4)$$

where the specific heat ($C_P$) values were taken from Refs. 47 and 49. It is worth noting that in certain cases where the specific heat depends strongly on the magnetic field, Eq. (4) is may be not suitable for the calculation of the adiabatic temperature change $\Delta T_{ad}$. However, according to a recently reported work by Song et al [47], it can be clearly seen that for h-ErMnO$_3$, the specific heat and particularly the ratio ($C_p/T$) at temperatures far above the ordering point of $Er^{+3}$ magnetic moments, depend slightly on the magnetic field, which enables us to reasonably evaluate the adiabatic temperature change in this temperature range using Eq (4). For practical reasons, the h-ErMnO$_3$ adiabatic temperature change is evaluated close to 20 K that is the liquid hydrogen temperature. Around this temperature, the ratio $C_p/T$ is about 1.35 J/kg K$^2$ whatever the value of the applied magnetic field [47]. The field dependence of $\Delta T_{ad}$ along the ab-plane is reported in Figure 8-b. For magnetic field variations of 2, 5 and 7 T within the ab-plane, the $\Delta T_{ad}$ at 19 K reaches values of about 1.5, 7.5 and 12 K, respectively. However, the $\Delta T_{ad}$ shown by h-ErMnO$_3$ crystals is expected to be much larger at low temperatures since the MCE usually maximizes around the magnetic ordering point. A detailed study of $\Delta T_{ad}$ in h-ErMnO$_3$ crystals using specific heat measurements under magnetic fields will be reported in the future.

More recently, the magnetocaloric properties of h-ErMnO$_3$ polycrystalline samples were investigated by Sattibabu et al [19] and Vinod et al [20]. In both works, the reported entropy change is much lower than for our single crystals. Under a magnetic field variation from 0 to 5 T, the maximum $\Delta S$ that can be reached by using h-ErMnO$_3$ polycrystals is only 14 J/kg K instead about 20.5 J/kg K shown by h-ErMnO$_3$ single crystals along the ab-plane (see Fig.5-a.). This result points out the negative impact of the magnetocrystalline anisotropy usually exhibited by non-cubic materials such as h-ErMnO$_3$ on the polycrystalline magnetocaloric performance [62]. As can be clearly seen in Figure 3-a, the h-ErMnO$_3$ compound presents a large magnetic anisotropy. Because of the random orientation of grains in polycrystalline samples, the obtained magnetization and accordingly the MCE under a given magnetic field are rather equivalent to the mean value of those associated to the main crystallographic axes (hard, intermediate and easy-axes). In the case of our crystals, the mean value of its entropy changes along the ab and c-orientations (15 J/kg K for 5 T) is practically similar to the reported entropy change for polycrystalline samples [19, 20]. Therefore, in magnetocaloric regenerators, it is necessary to orient the easy-axis of polycrystalline refrigerants parallel to the magnetic field direction. In this context, textured magnetocaloric powders under magnetic fields would be a promising solution [63]. It would enable us to maximize the MCE in the regenerators, increasing then the efficiency of magnetic cooling systems.

As demonstrated above, the h-ErMnO$_3$ single crystal generates a giant conventional MCE that can be obtained by the standard magnetization-demagnetization process (field variation), within the ab-plane. However, the large anisotropy shown by the MCE between the ab and c-orientations indicates that additional MCEs can be obtained by spinning the h-ErMnO$_3$ crystals around their a or b-axes in constant magnetic fields. Such process is of great interest from a practical point of view, since the resulting RMCE would enable the implementation of more compact and efficient rotary magnetic refrigerators with simplified designs [11, 27]. It is worth noting that the difficulty associated with the growth of crystals remains a serious obstacle to their utilization in RMCE-based cooling devices. However, this drawback can be avoided by using, for example, textured polycrystalline powders of these crystals under magnetic fields, which is more interesting from both practical and economical points of view. In this way, textured powders of anisotropic magnetic compounds using for example the described process in Ref. 63 would also enables us to generate large RMCEs without the need to single crystals.

The isothermal entropy change arising from the rotation of h-ErMnO$_3$ single crystals within the bc or ac planes was evaluated based on the entropy changes reported above. Considering initially the magnetic field parallel to the c-axis, the rotation of h-ErMnO$_3$ around the b (or a)-axis by an angle of 90 ° results in a change of the total entropy by [1, 14-16, 24, 27]

$$\Delta S_{c-ab} = \Delta S\ (H//ab) - \Delta S\ (H//c) \qquad (5)$$

with ΔS (H//ab) and ΔS (H//c) are the entropy changes associated with the magnetization of h-ErMnO$_3$ along the ab-plane and the c-axis, respectively. The temperature dependence of $\Delta S_{R,c-ab}$ for several constant magnetic fields is plotted in Figure 9-a. The associated adiabatic temperature change $\Delta T_{R,c-ab}$ that can be calculated by using Eq. (4) is reported in Figure 9-b as a function of magnetic field at 19 K. As shown, the h-ErMnO$_3$ compound provides a giant RMCE over the temperature range around 10 K. Surprisingly, the displayed RMCE is about three times larger than that shown by h-HoMnO$_3$ crystals [16] in a similar working temperature range (see Fig. 10-a). The rotation of h-ErMnO$_3$ crystal around the a (or b) -axis in constant magnetic fields of 2, 5, and 7 T leads to maximum entropy changes of 7, 17 and 20 J/kg K, respectively. Also, the generated $\Delta S_{R,c-ab}$ is even larger than that found in the vast majority of orthorhombic RMn$_2$O$_5$ and RMnO$_3$ phases, usually known for their gigantic magnetocrystalline anisotropy [27] such as TbMnO$_3$ (18 J/kg K for 7 T) [13, 14, 24], o-DyMnO$_3$ (18 J/kg K for 7 T ) [18], HoMn$_2$O$_5$ (12.3 J/kg K for 7 T) [11] and TbMn$_2$O$_5$ (13.35 J/kg K for 7 T) [12]. On the other hand, in a constant magnetic field of 7 T initially parallel to the c-axis, the adiabatic temperature change resulting from the rotation of h-ErMnO$_3$ crystals at 19 K between their c and a (or b)-axes is about 12 K. Because of the reversible character of the adiabatic temperature change (absence of hysteresis), the rotation of h-ErMnO$_3$ crystals in the opposite direction at 31 K which is closer to the critical (boiling) point of hydrogen (33 K), enables us to reduce their temperature down to about 20 K which is the temperature of fully liquid state of hydrogen. This generated negative thermal effect could be then used for example to liquify gaseous hydrogen [8] as proposed in Ref.11.

The giant RMCE observed in h-ErMnO$_3$ single crystals is mainly attributed to their unique magnetic features when compared to other hexagonal RMnO$_3$ phases [27]. As shown in figure 3-b, the first-order metamagnetic transition (involving 2a sites) giving rise to a large magnetocaloric effect along the c-axis, occurs at very low temperatures (around 3 K). Consequently, the MCE along this direction is peaked around 3 K and rapidly declines above this temperature (Fig. 8-a) due to the weakness of AF-F transition. In contrast, because of the "paramagnetic"-like behaviour of Er$^{3+}$ magnetic moments along the ab-direction, the MCE peak position increases significantly with magnetic field change. When the latter is varied from 2 to 7 T, the MCE maximum is shifted from 3.5 to 9.5 K for H//ab and, only from 2.7 to 3.7 K for H//C (Fig. 5). Consequently, the MCE anisotropy is maximized between the c and ab-orientations particularly under relatively high magnetic fields, leading then to a giant RMCE. This markedly contrasts to other hexagonal $RMnO_3$ single crystals [16, 26, 27]. For example, under a magnetic field changing from 0 to 7 T at 11 K, the entropy change shown by h-$ErMnO_3$ along the ab-plane is about 10 times larger than that along the c-axis. In similar conditions of temperature and magnetic field, the ΔS shown by h-$HoMnO_3$ along the ab-plane is only 1.4 times larger if compared with its equivalent following the c-axis [16].

The complexity as well as the role of the magnetocrystalline anisotropy in the determination of the RMCE in h-$ErMnO_3$ crystals was also investigated. For this purpose, the coherent rotational (CR) model [11-12, 14, 24, 52] was used to calculate the entropy change $\Delta S_{MCA}$ associated with their rotation between the c and ab directions. This particularly allows us to quickly figure out whether the RMCE arises directly from the magnetocrystalline anisotropy, while informing us on its complexity. In Figure 10-b, we present the calculated $\Delta S_{MCA}$ as a function of angle ß at 8 and 15 K for a constant magnetic field of 7 T. β represents the angle between the magnetic field direction and the easy-axis (a or b). When the h-$ErMnO_3$ crystal is rotated from β = 90 ° to β = 0 ° under 7 T, the associated $\Delta S_{MCA}$ is 16.5 J/kg K and 8 J/kg K at 8 K and 15 K, respectively. The corresponding experimental values are 16 and 16.3 J/kg K, respectively. These findings demonstrate that just below the ordering point of $Er^{3+}$ magnetic moments along the ab-plane (~ 11 K, under 7 T), the experimental and calculated values are in fair agreement. This underlines the fact that the shown RMCE at this temperature is mainly arising from the magnetocrystalline anisotropy. In contrast, a marked deviation is observed between the calculated and experimental rotating entropy changes at 15 K. Such difference is probably due to the fact that the CR model does not include thermal fluctuations contributions usually taking place at high temperatures. At temperatures below 8 K, it is highly challenging to identify the mechanism behind the RMCE because of the complexity of the magnetocrystalline anisotropy. In this temperature range, the magnetization presents a pronounced jump after crossing a critical magnetic field making more difficult the determination of anisotropy constants by using the CR model, pointing out to the limitations of the latter.

The last point in this paper concerns the efficiency of h-$ErMnO_3$ single crystals in functional magnetocaloric devices. For this purpose, the suggested parameter in Refs. 64 and 65 and given by $\eta = |Q/W|$ is utilized as a figure of merit. Q is the thermal energy generated by the magnetocaloric material close to the transition temperature under external magnetic fields, while W represents the magnetic work that is required to drive this isothermal heat. Following the reported procedure in Moya et al [64], both Q and W can be determined from entropy change curves and magnetic isotherms, respectively. Preliminary investigations reveal that all the quantities $\eta$, Q, and W depend strongly on the crystal orientation, magnetic field and initial temperature. For example, when the used magnetic field varies from 0 to 2 T within the ab-plane at ~3.5 K, a magnetic work of 131 J/kg is needed to produce an isothermal heat of 43.4 J/kg, leading then to an efficiency of about 0.33. For a magnetic field of 7 T, the required magnetic work and the generated heat (at 11 K) markedly increases to reach 683 J/kg and 247 J/kg, respectively. The associated efficiency is slightly increased to 0.36 which can be more probably explained by the saturation of magnetization along the ab-direction. On the other hand, it was observed that under relatively low magnetic fields reachable by permanent magnets, the magnetocaloric response of h-ErMnO$_3$ is significantly efficient along the c-direction. For a driving magnetic field of 2 T applied along the c-axis at ~2.75 K, it was found to be 0.55 (W = 28 J/kg, Q = 15.52 J/kg), instead only 0.33 for the ab-plane (at ~3.5 K). It is worth noting that the $\eta$ parameter is used as a guide to identify the efficiency of magnetocaloric materials only in isothermal processes and not within a complete thermodynamic cycle. In fact, the thermodynamic efficiency of magnetocaloric devices depends on many other parameters such as the temperature gradient that a material can achieve after several AMR-cycles [1], thermal losses and additional works caused by eddy currents, magnetic field and energy recovery via the compensation of magnetic forces [66].

## IV. Conclusions

In this work, we have mainly focused on the magnetic and magnetocaloric properties of h-ErMnO$_3$ single crystals. The latter reveal a giant conventional magnetocaloric effect that can be generated by applying magnetic fields along the ab-plane. In the magnetic field change of 7 T applied within the ab-plane, the resulting isothermal entropy change presents a maximum value of 22.7 J/kg K close to 10 K. In contrast, the partial polarization of Er$^{3+}$ magnetic moments (located at 2a sites) along the c-axis via a field-induced magnetic transition leads to a less-important MCE that is peaked within a narrow temperature range with a maximum of 11.4 J/kg K (7 T) close to 3 K. Additionally, the distinct features of Er$^{3+}$ magnetic moments along the c-axis and the ab-plane lead to a strong anisotropy of the MCE around 10 K. Consequently, the h-ErMnO$_3$ compound presents a giant RMCE that is several times larger than that showed by other hexagonal RMnO$_3$ phases. The rotation of h-ErMnO$_3$ single crystals in a constant magnetic field of 7 T within the ac or bc-planes enables to generate a maximum entropy change of 20 J/kg K. These findings open the way for the implementation of h-ErMnO$_3$ compounds as refrigerants in cryo-magnetocaloric liquefiers. On the other hand, the application of the coherent rotation model unveiled the complexity of the h-ErMnO$_3$ magnetocrystalline anisotropy. Preliminary calculations show that at temperatures just below the ordering point of Er$^{3+}$ magnetic moments (under 7 T) within the ab-plane, roughly T = 8 K, the RMCE is mainly contributed from the magnetocrystalline anisotropy. However, for temperatures far above and below 8 K, the contribution to the RMCE of thermal fluctuations as well as metamagnetic transitions occurring along the c-axis must be considered, respectively.

**Acknowledgments**

The authors thank M. Castonguay, S. Pelletier and B. Rivard for technical support. We acknowledge the financial support from NSERC (Canada), FQRNT (Québec), CFI, CIFAR, Canada First Research Excellence Fund (Apogée Canada) and the Université de Sherbrooke.

**Figure captions**

**Figure 1:** (left) Micro-Raman spectra at 10 K for h-ErMnO$_3$. Those of YMnO$_3$ considered as a reference for hexagonal RMnO$_3$ phases are also reported for comparison. The backscattering and light polarization configurations b(cc)b and c(ab)c, allow the observation of A$_1$ and E$_2$ phonon symmetries, respectively. (right) Crystallographic structure of h-ErMnO$_3$. R$_1$ and R$_2$ correspond to 2a and 4b sites, respectively.

**Figure 2:** (a) Temperature dependence of ZFC and FC magnetizations of h-ErMnO$_3$ under a magnetic field of 0.1 T along the ab-plane and the c-axis. (b) Thermomagnetic curves of h-ErMnO$_3$ under magnetic fields going from 0.2 T up to 4 T (H // c).

**Figure 3:** (a) Isothermal magnetization curves of the single crystal h-ErMnO$_3$ at 2 K along the ab-plane and the c-axis. (b) Collected magnetic isotherms along the c-axis at very low temperatures and, under low magnetic fields.

**Figure 4:** Isothermal magnetization curves of the single crystal h-ErMnO$_3$ for (a) H//*ab*, (b) H//*c and* (c) H//c.

**Figure 5:** Temperature dependence of the isothermal entropy change in h-ErMnO$_3$ under different magnetic fields for (a) H//*ab* and (b) H//*c*.

**Figure 6:** (a) Temperature dependence of the critical field B$_C$ (along the c-axis) corresponding to the metamagnetic transition taking place in the single crystal h-ErMnO$_3$. (b) The associated entropy change deduced from Clausius–Clapeyron equation.

**Figure 7:** Temperature dependence of the Raman-active phonons E$_2$(5) (a) and A$_1$(9) (b) for both h-ErMnO$_3$ and YMnO$_3$ single crystals. (Right) Description of ion relative motions.

**Figure 8** (a) Comparison between the h-ErMnO$_3$ entropy changes related to H//*ab* and H//*c* for 7 T. (b) Adiabatic temperature change of h-ErMnO$_3$ as a function of magnetic field at 19 K applied within the ab-plane. (c) and (d) Arrott plots of h-ErMnO$_3$ close to the ordering point of Er$^{3+}$ magnetic moments along the c-axis and the ab-plane, respectively.

**Figure 9:** (a) Entropy changes related to the rotation of h-ErMnO$_3$ from the *c* to the *a (or b)* direction by 90 ° in different constant magnetic fields, with magnetic field initially parallel to the *c*-axis. (b) Associated adiabatic temperature change as a function of magnetic field at 19 K.

**Figure 10:** (a) Temperature dependence of the rotating entropy change in the hexagonal ErMnO$_3$ and HoMnO$_3$ [16] single crystals for 7 T. (b)The calculated rotating entropy change from the coherent rotational model for h-ErMnO$_3$. The magnetic field is initially parallel to the c-axis. β is the angle between the easy direction (a or b axes) and the magnetic field.

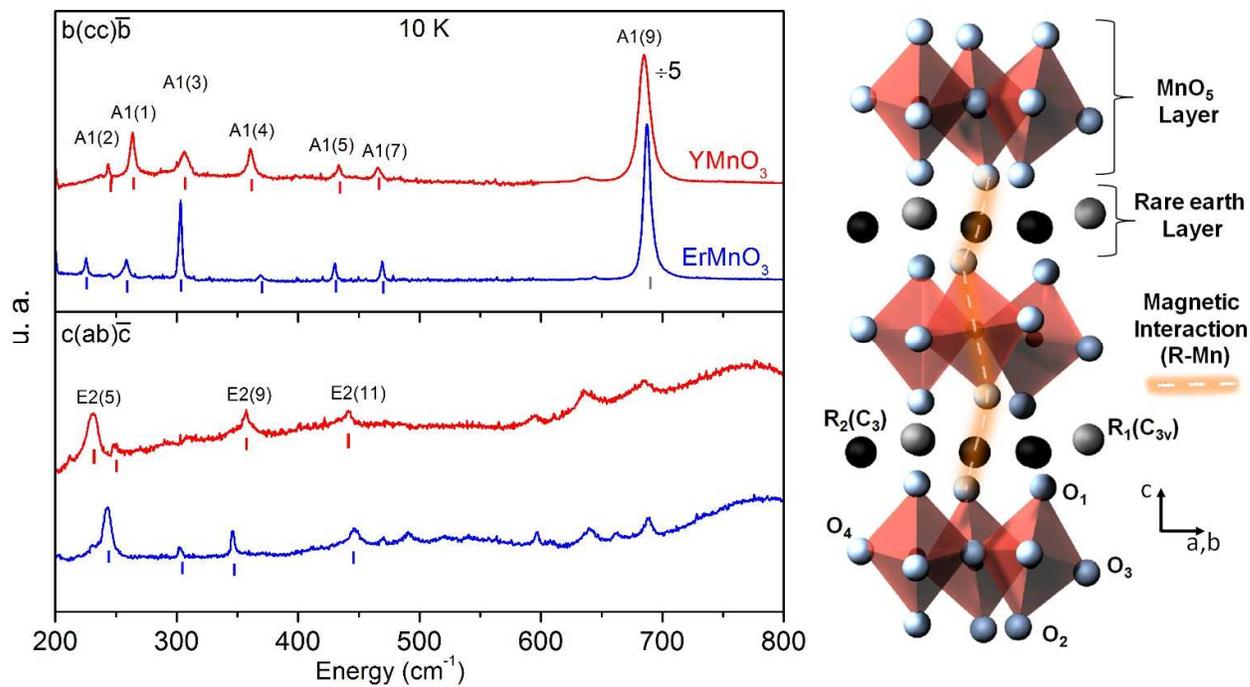

Figure 1

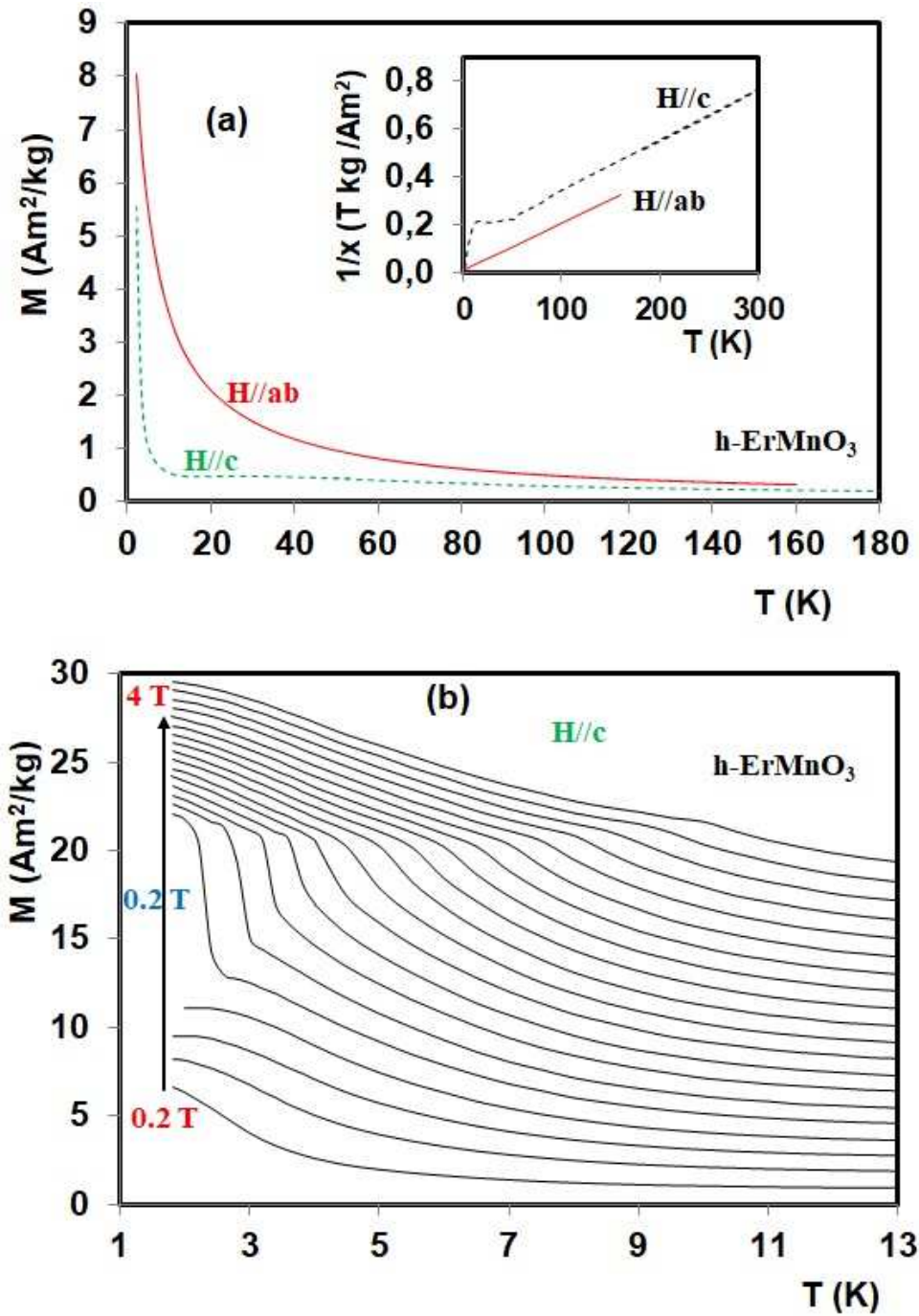

Figure 2

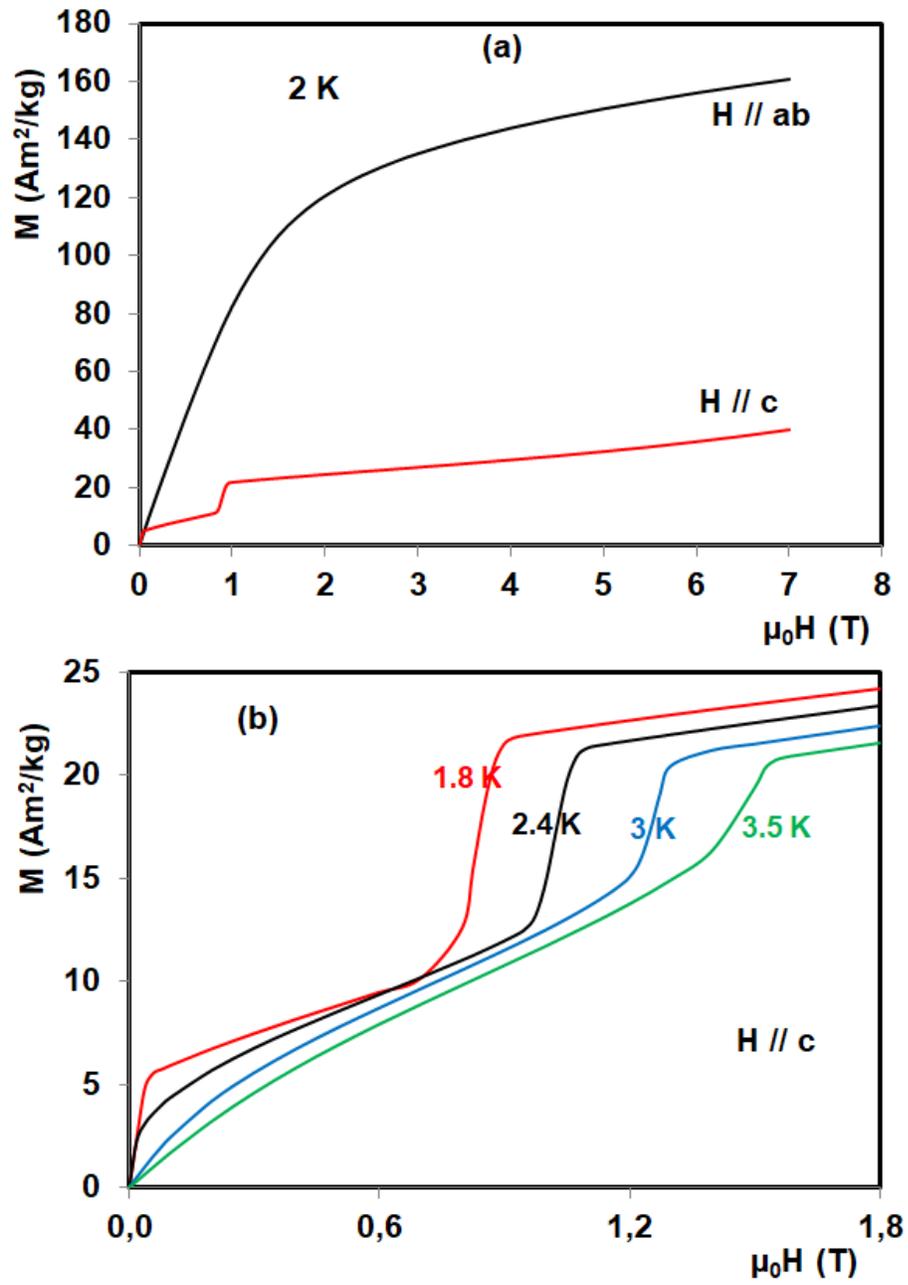

Figure 3

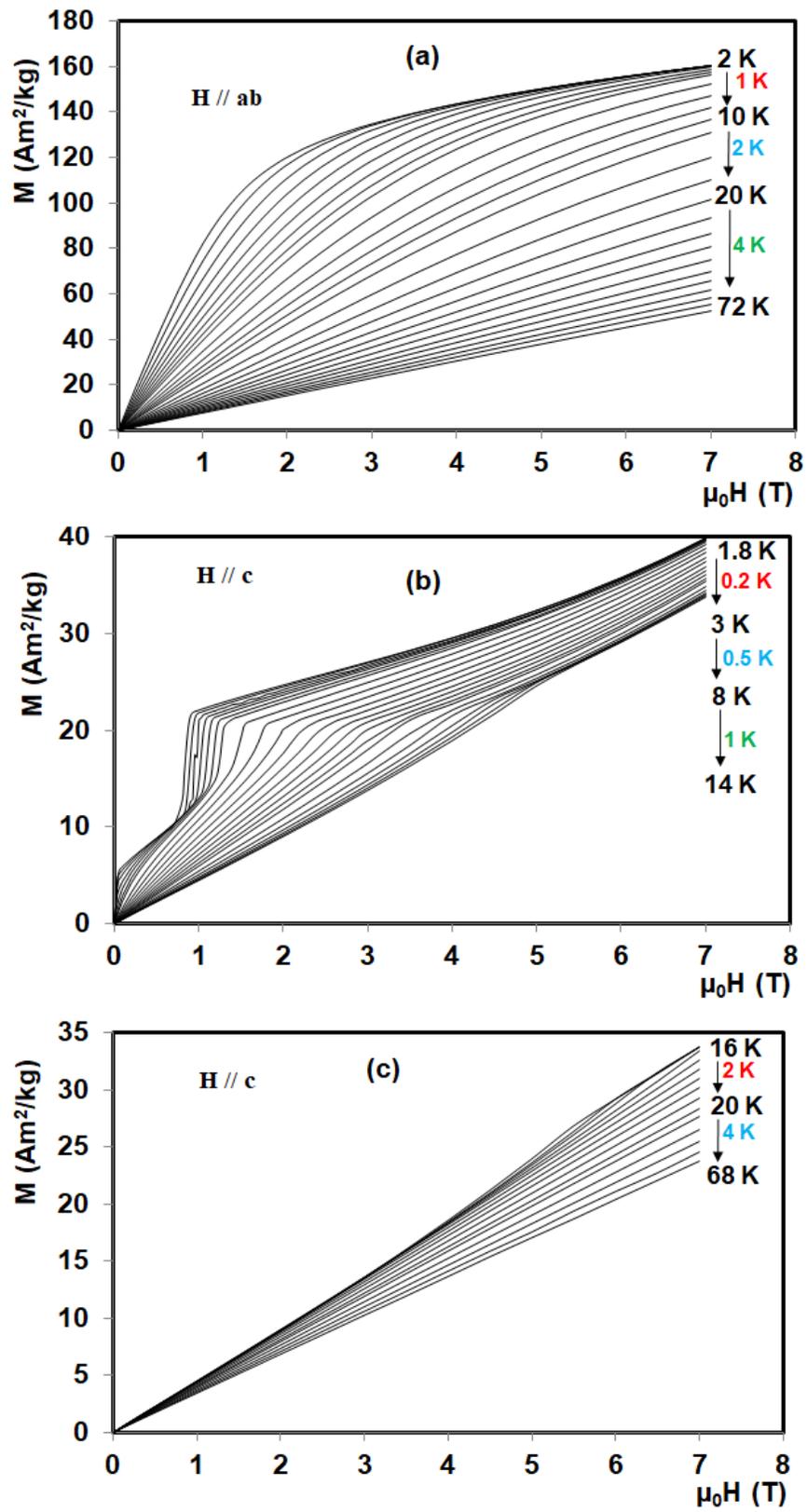

Figure 4

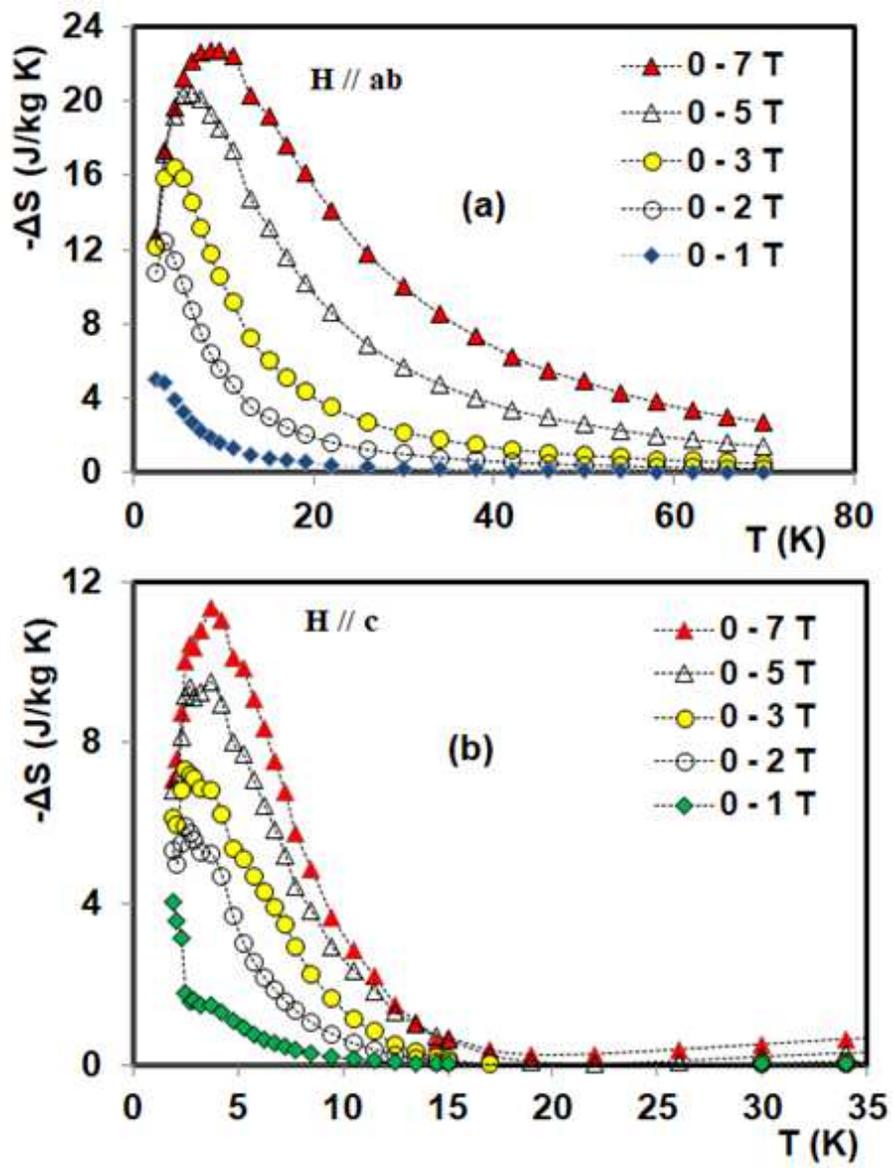

Figure 5

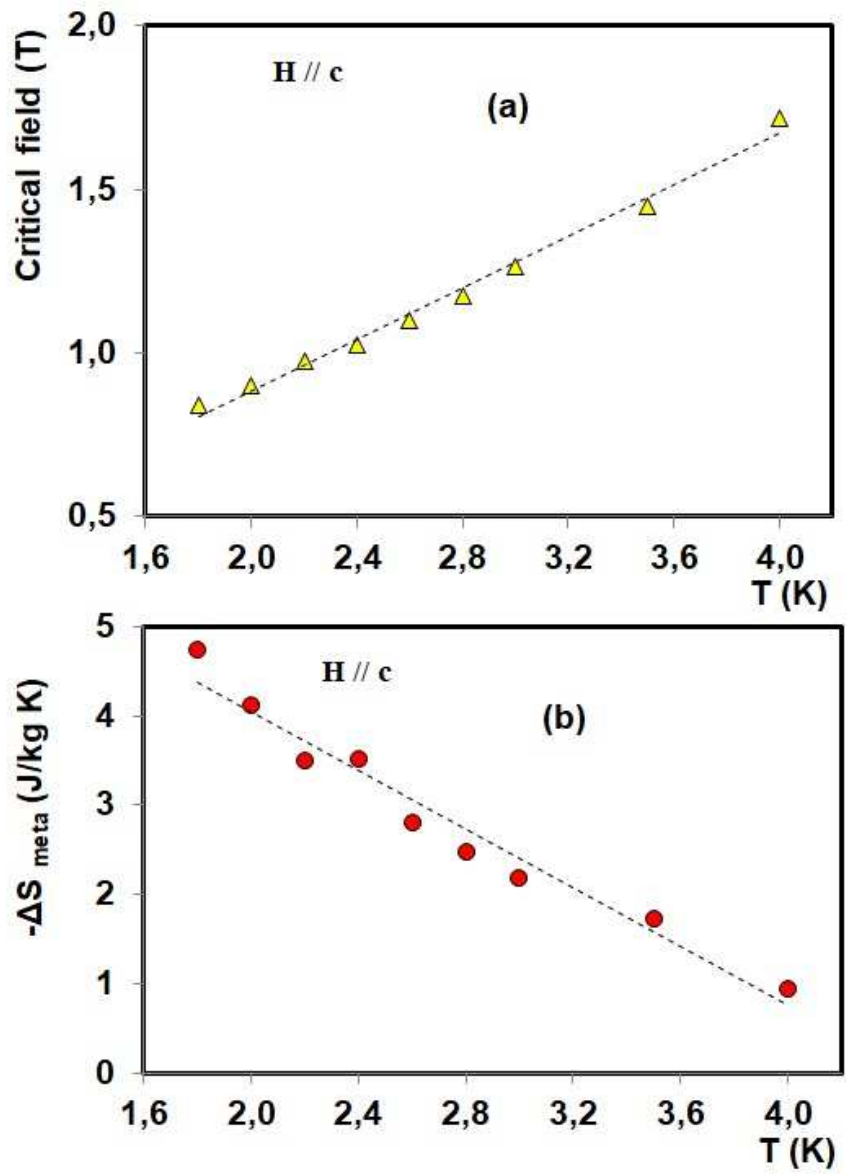

Figure 6

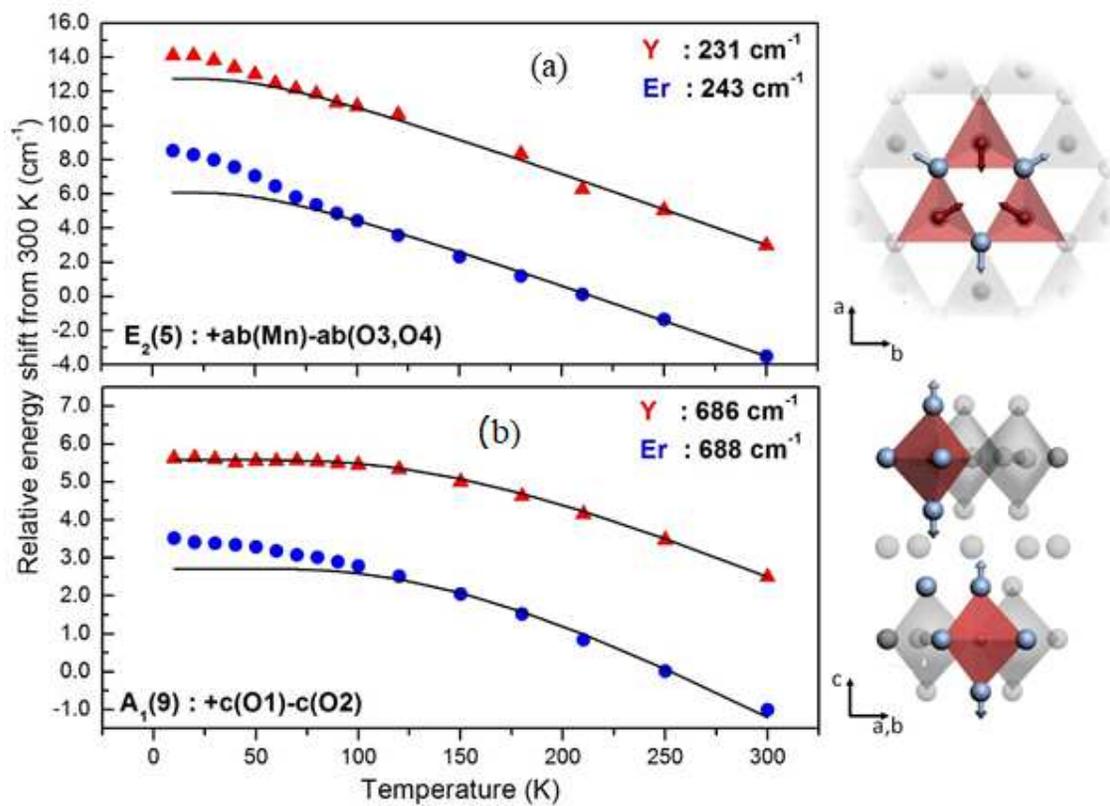

Figure 7

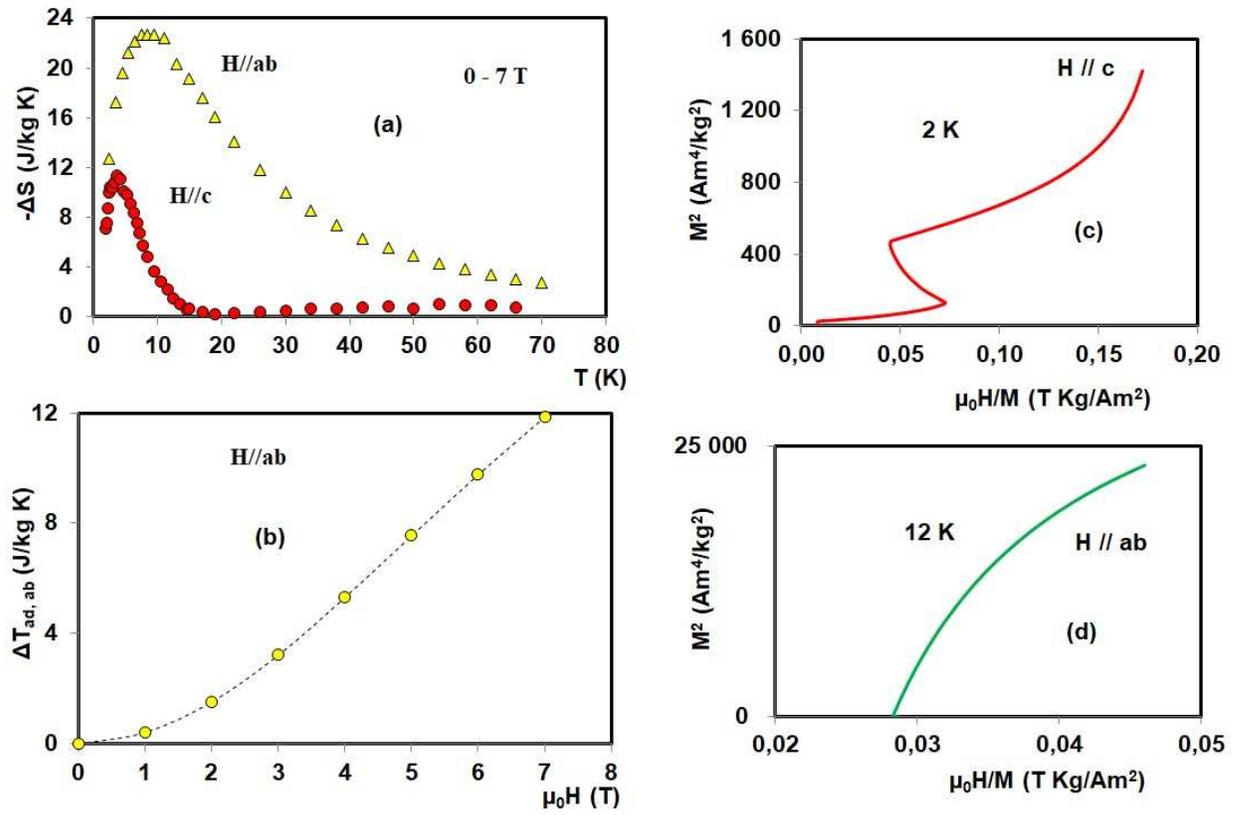

Figure 8

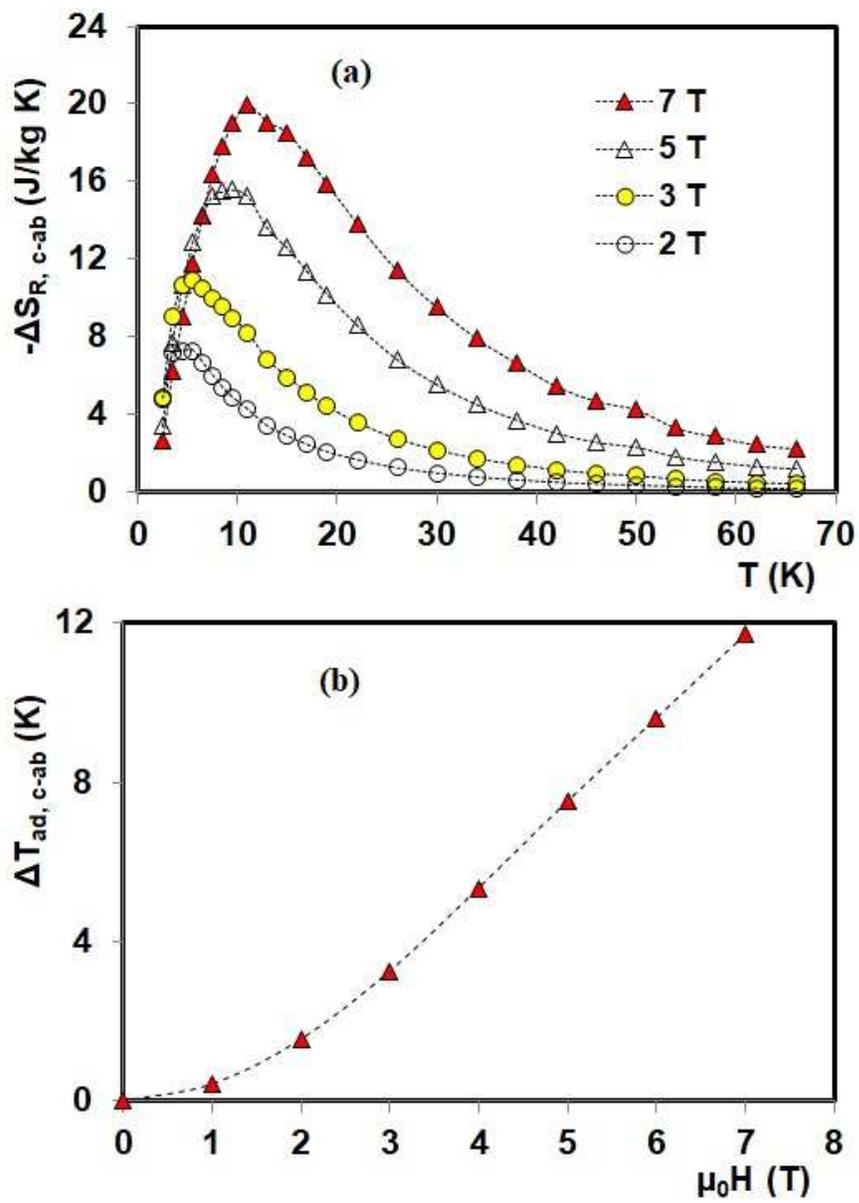

Figure 9

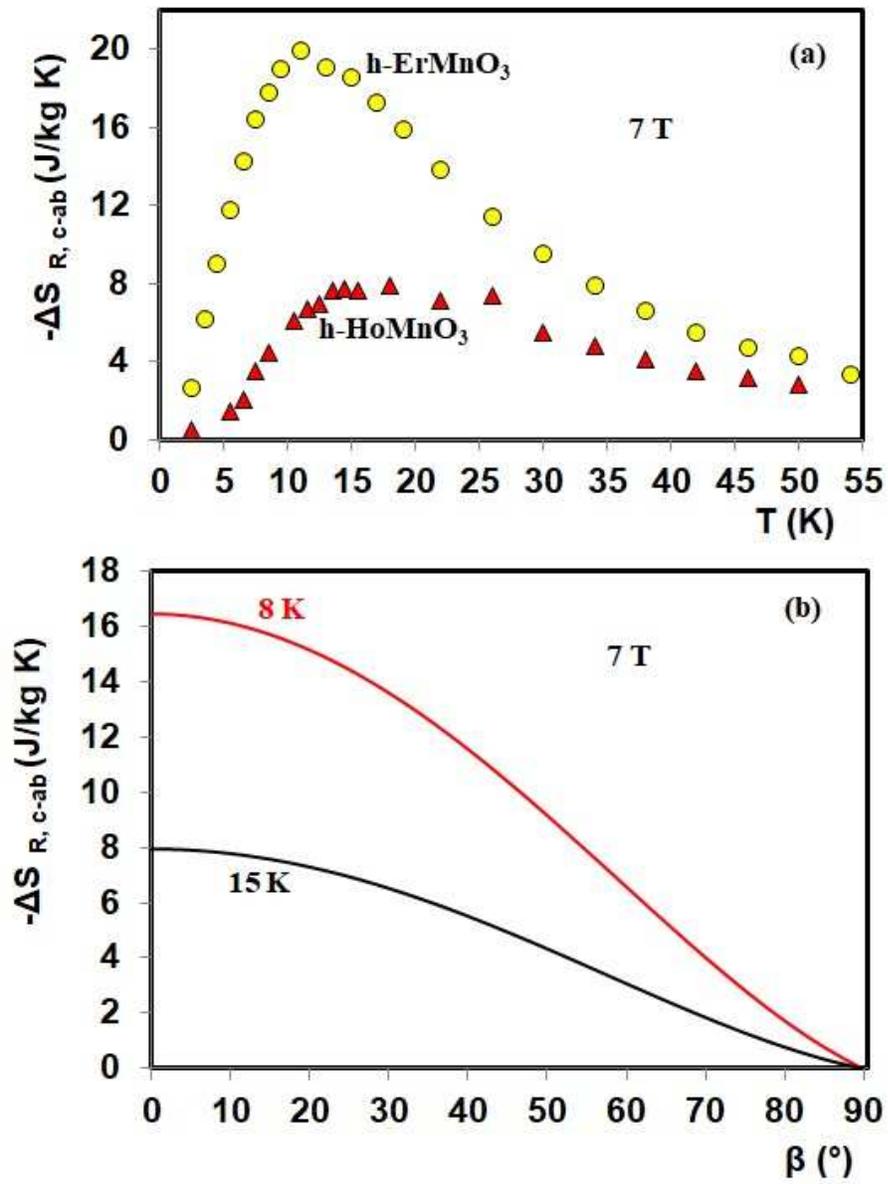

Figure 10